\documentstyle[12pt,aps,epsf]{revtex}

\begin{document}

%\hfill {}

%\hfill {}

\vskip 1cm

\title{
Gluonic states in two space dimensions.
}

\vskip 1cm
\author{Gabriel Karl\footnote{permanent address: Dept of Physics, University of Guelph, Ontario, Canada} and Jack Paton}
\address{Department of Physics (Theoretical), 1 Keble Road , Oxford, United Kingdom OX1 3NP}

%j.paton1@physics.ox.ac.uk

\date{August 14, 1999}
\maketitle

\vspace{0.5cm}
\begin{center}
{\bf Abstract}
\end{center}
\vspace {.2cm}

\begin{abstract}
We discuss the ``spectroscopy'' of gluonic states in systems with two space dimensions, using simple models to mimic the results of lattice gauge theory computations. We first discuss the quantum numbers of these systems, including charge conjugation. Two types of systems are discussed in detail: ``gluelumps'' which have a heavy adjoint color charge at the origin and glueballs which are composed entirely of glue. Both systems are discussed using the bag model and the flux-tube model. For glueballs the model spectra are compared with the results of Teper. Both models capture many features of the numerical results. 

\end{abstract}

\newpage

\section{Introduction}
The numerical results of lattice Monte Carlo computations are becoming increasingly reliable. This presents the challenge of understanding these results and the physics underlying them. 
In this  endeavour simple models are still very useful \cite{Paton}. The use of lattice simulations  allows the study of systems with varying numbers of colors and even different numbers of space dimensions \cite{Teper} yielding more detailed tests of models. In this paper we adapt simple models of hadron spectroscopy to such systems in the hope of understanding the lattice results. Earlier \cite{gluelump} we have used the bag model to understand the results of Michael \cite{Michael} for a ``gluelump'' system in three space dimensions. Here we use the bag model and flux-tube model to understand the ``spectroscopy'' of 
\begin{itemize}
\item gluelumps in two space dimensions, where lattice results do not yet exist, and 
\item glueballs in two space dimensions, where a detailed numerical study on the lattice was published by Teper \cite{Teper}.
\end{itemize}
The lay-out of our paper is as follows: In Section II we discuss the quantum numbers of states in two plus one dimensional chromodynamics, and in particular the charge conjugation quantum number of pure glue states. In Section III we discuss a bag model for gluelumps in two space dimensions. In these systems there is a heavy color charge at the origin which is neutralized by one or more gluons. This system is the simplest one to treat in the bag model because of the absence of spurious states associated with centre of mass motion. Unfortunately there are as yet no lattice data with which we can compare our results. In the next section (IV) we discuss models of glueballs in two space dimensions. Here the problem of spurious states in the bag model is dealt with by using a harmonic oscillator approximation. We also treat this system in the flux-tube model (which has already been done in \cite{Moretto}, \cite{Johnson}, \cite{Paton}) in a particularly transparent version of this model in which mases are given analytically in terms of a system of oscillators. We also in the final part of  Section IV extend the model to give a prediction for the spectrum of gluelumps in two space dimensions. In Section V we compare the model spectra with each other and with the lattice results of Teper {\cite{Teper}}, and draw conclusions.   \\
\section{Quantum Numbers of Pure Glue Theory States in Two  Space Dimensions.} 
We first make a couple of general points about the multiplet structure and quantum numbers of  ``spectroscopy'' in two space dimensions for an arbitrary number $N_C$ of colors:
\begin{itemize}
\item As is well known, the irreducible  representations of the two dimensional rotational group are one dimensional, characterized by the integer angular momentum quantum number, $m$. Pure gauge theory is also invariant under the parity operator, $P:(x,y)\rightarrow(x, -y)$ which transforms $m$ to $-m$, giving two dimensional multiplets in all cases except $m=0$. For $m=0$ the state is also characterized by its eigenvalue of $P$. 
\item In addition there is invariance under a suitably defined ``charge conjugation'' operation  which changes a gluon field variable to its complex conjugate. If we use the real form of the $N^2-1$ dimensional adjoint gluon representation each of the field variables will have $C=\pm 1$.
\begin{enumerate}
\item The case of SU(2) gauge theory is special and we treat it first. In terms of real fields $g_1$, $g_2$ and $g_3$, we require $C$ to be such that the three gluon coupling, with color dependence $\epsilon_{ijk}g_i g_j g_k$, be invariant. (If this is the case  the four gluon coupling will be invariant automatically.) For non-trivial $C$ we therefore require two of the fields to have negative $C$ and the third positive. The conventional choice would be to take $g_1$ and $g_3$ with $C$ negative so that 
\begin{equation} C[g_1 + i g_2, g_1 - i g_2, g_3] = -[g_1 - i g_2, g_1 + i g_2, g_3].\end{equation}
and we can say loosely that the gluon has negative $C$, though strictly speaking this applies to $g_1$ and $g_3$ only, with the $C$ of $g_2$ being positive.\\
As is well known, there exists for SU(2) (but not for higher SU(N)) an operator related to $C$ and similar to the G-parity of hadron flavour physics,  of which all members of the gluon multiplet are eigenfunctions with the same eigenvalue. This is the operator
\[ G = C \exp(i\pi T_2)  \]
where $[T_1, T_2, T_3]$ are the generators of SU(2). We have 
\[ G[g_1,g_2, g_3] =  + [g_1,g_2, g_3]  \]
so that there is no selection rule analogous to the hadron physics selection rule that allows only an even number of pions (flavor G-parity  $-1$) to couple together. This is important to allow three gluon coupling.\\
We can show immediately from the above that any local gauge invariant operator, which must necessarily be a color singlet combination of the gluon fields and their derivatives at a point, must  have $C=+1$. This immediately follows from the fact that $C=G$ for such a field and $G$ is necessarily $+1$. It follows that the physical states (created from the vacuum by local color singlet operators) must necessarily have $C=+1$ in the case of pure SU(2) gauge theory (in any number of dimensions and  in the absence of sources). 
\item In the case of three colors, SU(3), again using the real form of the adjoint representation we require that $C(g_i) C(g_j) C(g_k) = +1$ for all sets $(i,j,k)$ for which the structure constants $f_{ijk}$ are non-zero. Making the conventional choice of $g_3$ and $g_8$ as the Abelian generators, these must have negative $C$ and we can, analogously to equation [1] take
\[ 
C(g_1 +ig_2, g_1-ig_2,g_3, g_4 +ig_5, g_4-ig_5, g_6+ig_7, g_6-ig_7,g_8) \] \begin{equation} = -(g_1 -ig_2, g_1+ig_2,g_3, g_4 -ig_5, g_4+ig_5, g_6-ig_7, g_6+ig_7,g_8)  \end{equation}
Since there is no analogue of the $G$-operator of SU(2) we can no longer conclude that physical pure glue states must necessarily have positive $C$, and in fact both our models as well as the lattice Monte Carlo calculation  exhibit states of both positive and negative $C$.
\item The case of general SU(N) is analogous to the case $N=3$ above. The  operator $C$ applied to the glue field operators corresponding to all Abelian generators gives a negative sign and $C$ applied to any raising (lowering) operators  gives the corresponding lowering (raising) operator with a minus sign.
\end{enumerate}

\end{itemize}

\section{Bag model of two dimensional gluelumps}
We define the bag model in two space dimensions exactly analogously to the definition in three space dimensions. The ``gluelump'' consists of an adjoint source at the origin and in the model we assume it is surrounded by a circular region in which one or more approximately free gluons  exist to neutralize the color.  As usual in the bag model we use color Coulomb gauge, first treat non-interacting gluons confined to the bag, and then treat the effect of the instantaneous color Coulomb interaction as a perturbation. 
\subsection{Gluon modes in the bag}
The fields of gluon modes confined to a circular bag of radius $R$ obey Maxwell's equations in $2+1$ dimensions which are identical to the full Maxwell's equations in $x$ and $y$  with ${\bf B}$ in the $z$-direction, ${\bf E}$ in the $xy$ plane and ${\bf E}$ and $B$ independent of $z$. Thus the ${B}$-field is a scalar and the ${\bf E}$-field is a two component vector in the two space dimensions.\\
Modes of definite frequency $k$ are given by ($r,\phi$ are plane polar coordinates):
\[ B= \exp (i m \phi)\; J_m(kr);\:\: {\bf E}=\frac{1}{ik}\left(\frac{\partial}{-\partial y},\frac{\partial}{\partial x}\right)B    \]
where the mode frequency $k$ is determined by the boundary  conditions equivalent to the three dimensional boundary conditions that ${\bf r}.{\bf E}$  and ${\bf r \times B}$ vanish at the bag radius $R$. ($J_m(x)$ is a Bessel function). The first of these conditions is automatic and the second implies that $B = 0$ at $r=R$. Thus we have  the single condition that  $J_m(\beta_{mi})=0$ where $\beta_{mi}$ is the $i$th zero of $J_m(x)$ and $k=\beta/R$. The lowest values of $\beta$ for $m=0,1...4$ are:
\[m=0:\:\:\: 2.4048, 5.5201,  8.6537,  11.7915 \]
\[m=1:\:\:\: 3.8317, 7.0156, 10.1735 \]
\[m=2:\:\:\: 5.1356, 8.4172, 11.61988 \]
\[m=3:\:\:\: 6.3802, 9.7610 \]
\[m=4:\:\:\: 7.5883,11.0649 \]
and ignoring for the moment the color Coulomb interaction between gluon and source the one gluon bag energies and radii are determined by minimizing
\[ E = \frac{\beta}{R} + \Lambda \pi R^2  \]
where $\Lambda$ which has dimension $L^{-3}$ is the two dimensional  bag constant and $\hbar=c=1$. We get 
\[ R=\left(\frac{\beta}{2 \pi \Lambda} \right)^{\frac{1}{3}}; \]   \begin{equation} E_{\rm min}= 3\left(\frac{\beta^2 \pi \Lambda}{4}\right)^\frac{1}{3}   \end{equation}
Following the discussion of Section II above, we label the states $|m|^C$ for non-zero $m$, and $0^{PC}$ for $m$ zero, so that we obtain the ordering of the lowest one-gluon levels, ignoring the color Coulomb and other interactions, as:
\[ 0^{--}, 1^-, 2^-,0^{--}, 3^-, 1^-, 4^-  \]
with a comparatively small spacing between the $2^-$ and first excited $0^{--}$ states, as well as between the $3^-$ and first excited $1^-$ state.
In determining the overall $C$ of a state we assume here that the adjoint source has positive charge conjugation.\\
There may also be states in which the source is neutralized by  two or more gluons. The lowest such two gluon state will be made out of two ground state gluons. It has
 quantum numbers $m^P = 0^+$. However in SU(2) only the antisymmetric coupling of three adjoints exists and since this state is symmetric in space it is forbidden by Bose symmetry. In all the higher color groups it exists and has $C=+1$ with our convention on the $C$ of the source. In this approximation it lies just below the first $2^-$ one gluon state. The next two gluon state contains a ground state $0^{--}$ gluon and a $1^-$ gluon. In SU(2) it has $|m|^C =1^-$  and lies just below the lowest $3^-$ one gluon state. For the higher color groups it is degenerate with an $|m|^C =1^{+}$ state. In general   the lowest states of quantum numbers $|m|^{\pm}$ ($|m|^-$ only for $SU(2)$ ) may be made from one excited gluon and a ground state gluon. The lowest three gluon state  has quantum numbers  $0^{--}$, the same as the ground state, and lies just below the lowest $4^-$ one gluon state. The only quantum numbers which we have not yet generated are $0^{+-}$ and $0^{-+}$. The  lowest $0^{+-}$ state may be made from two $m=0$ gluons, one with $\beta=2.4048$ and one with $\beta=5.5201$ and is also approximately degenerate in energy with the lowest $4^-$ one gluon state.   The lowest  $0^{-+}$ state, possible except in the case of two colors, lies considerably higher in energy.   The level ordering for all these states is given in the second column of Table I, where the masses are given in terms of the string tension parameter $\sigma$. (See Subsection III C below.)
\subsection{Colour Coulomb interactions}
This ordering may be affected by the color Coulomb interaction of the gluon or gluons with the source as well as self-energies and, in the case of the two- and three-gluon states, by the interactions between the gluons. 
\subsubsection{One Gluon States}
We treat  first the case of a single gluon  plus source:
\\In two space dimensions the color Coulomb interaction  between two point adjoint charges takes the form (the value of the Casimir  operator for interaction of two adjoints in the singlet state being $N_C$, the number of colors):
\[ V(r) = N_c \alpha_s \ln(|{\bf r_1 - r_2}|/r_0)   \]
where   $\alpha_s={g_s}^2/2 \pi$ and $r_0$ is an arbitrary length parameter which just adds a constant to the potential. The total color Coulomb energy is independent of $r_0$ when self-energies are taken into account. To see this, note that the source self-energy (in electrostatics this is $\frac{1}{2}Q V_Q(0)$) depends on $r_0$ through $\frac{1}{2}N_c\alpha_s \ln r_0$, and the gluon self-energy, which is 
\[ -\frac{\int\int d^2{\bf r_1}d^2{\bf r_2} {\bf E^2(r_1) E^2(r_2)} \frac{1}{2}N_c \alpha_s \ln |{\bf r_1 - r_2}|/r_0}{\int\int d^2{\bf r_1}d^2{\bf r_2} {\bf E^2(r_1) E^2(r_2)}}  \]
also depends on $r_0$ through $\frac{1}{2}N_c\alpha_s \ln r_0$. Finally, the gluon -source mutual energy is
\[ \frac{\int d^2 {\bf r E^2(r)} N_c {\alpha}_s \ln r/r_0}{\int d^2 {\bf r E^2(r)}}  \]
which depends on $r_0$ through $-N_c {\alpha}_s \ln r_0$. Thus the terms in $\ln r_0$ cancel, as claimed.\\
In the above, we have used the unconfined two-dimensional Coulomb Green's function  \[ g({\bf r_1, r_2}) = \ln(|{\bf r_1 - r_2}|/r_0). \]. This may be expanded  as 
\[ \ln r_>/r_0 - \sum_{n=1}^{\infty} \frac{1}{n}\left(\frac {r_<}{r_>}\right)^n \cos n(\phi_1-\phi_2) \]
where $\phi_1$ and $\phi_2$ are the polar angles of ${\bf r_1}$ and ${\bf r_2}$ and $r_>$ and $r_<$ are the larger and smaller of the  radial variables.  The modification of the Green's function to take account of the confining boundary condition that the normal component of the net $\bf E-$field is zero on the bag boundary affects the terms  in the second sum only. So, neglecting the angular correlations in the gluon self-energy integral, we can replace the confining Green's function by $\ln r_>/r_0$. Since the result is independent of $r_0$, as explained above, we take it equal to the bag radius, $R$ so that the Green's function vanishes at $r=R$. This gives the following expression for the total color Coulomb energy, including self-energies (and ignoring an infinite piece in the source self-energy):
\[ N_c\alpha_s \left(\frac{1}{2} \ln R  + 
\frac{\int_0^R dr\: r {\bf <E^2(r)>} \ln(r/R)}{\int_0^R dr \:r {\bf <E^2(r)>}} - \frac{\int_0^R dr\: r {\bf <E^2(r)>} \ln(r/R) \int_0^r dr^{\prime} r^{\prime} {\bf <E^2(r^{\prime})>}}{ \left(\int_0^R dr r {\bf <E^2(r)>}\right)^2} \right)   \] 
\\ We now change variables from $r,r^{\prime}$ to the dimensionless variables $x=kr, x^{\prime}=kr^{\prime}$, so that integrals go up to the relevant value of $\beta$. Using the value averaged over angles
\[ <{\bf E^2(r)}> = \left ( \frac{\beta}{R} \right )^2 \left ( \left (\frac{nJ_n(x)}{x} \right )^2 +(J_n^{\prime})^2 \right )  \]
we get for the total Coulomb energy
\[ N_c \alpha_s (C + A - B)  \]
where
\[ C = \frac{1}{2} \ln R/\beta, \] 
\[ A = \frac{\int_0^{\beta} dx\:x \ln x \left (\left (\frac{nJ_n(x)}{x} \right)^2 +(J_n^{\prime})^2 \right)} 
{\int_0^{\beta} dx\:x  \left(\left(\frac{nJ_n(x)}{x}\right)^2 +(J_n^{\prime})^2 \right)}, \] 
\[ B = \frac{\int_0^{\beta} dx\:x \ln x \left (\left (\frac{nJ_n(x)}{x} \right)^2 +(J_n^{\prime})^2 \right)\int_0^{x} dx\:x  \left (\left (\frac{nJ_n(x)}{x} \right)^2 +(J_n^{\prime})^2 \right)}{\left( {\int_0^{\beta} dx\:x  \left(\left(\frac{nJ_n(x)}{x}\right)^2 +(J_n^{\prime})^2 \right)} \right)^2}      \]
There still seems to be a residual dependence on scale since the $R/\beta = 1/k$ occuring in $C$  has dimensions. This arises because an infinite part of the source self energy  has been omitted. This will not affect our results for the spacing of the bag model levels, but only change their absolute values.  Unlike in three dimensions there is no natural way of normalising the two dimensional Coulomb potential.
\subsubsection{States of two and more gluons}
For a multi-gluon state with all gluons in the same spatial state it is not difficult to show that the color Coulomb perturbation is identical to that in the corresponding  single gluon state. For example, consider a two gluon state: The color Coulomb energy is given by (source self energy) + 2 (gluon self-energy) + 2 (gluon-source energy)  + (gluon-gluon interaction energy). However, the gluon-source energy is half the gluon-source energy in a one gluon state, since now the gluon and source couple to the adjoint representation. Also, the gluon-gluon interaction energy, which would be -2 times the gluon self-energy in two dimensional electrodynamics, contains a further factor of one half for the same reason. Thus, the color Coulomb energy of such a state is again given by the combination $N_C \alpha_s(A-B+C')$. Only one of the two-gluon states listed, that containing one gluon in the ground state and one in the first excited state requires a separate calculation of the color Coulomb correction to its energy. This is somewhat less than the Coulomb energy when both gluons are in the ground state. Because it turns out (see Section C below)  that the color Coulomb energies are always small, we have not listed those for two and three gluon states in the summary Table I.  

\subsection{Relations between parameters and bag model gluelump spectrum}
 In two space dimensions the squared coupling constant ${g_s}^2$ has dimension of mass. In order to find the  effect of the Coulomb interaction relative to the unperturbed levels in the bag, we need to relate the bag constant and ${g_s}^2$. In principle there should   be only one independent constant setting the scale of masses in the confining gauge theory.\\
We get an approximate relationship as follows: Teper \cite{Teper} finds  that in 2+1 dimensional SU($N_c$) lattice gauge theory 
the string tension $\sigma$ is related to the coupling $g$ by 
\[ \sqrt{\sigma} \approx 0.1975 N_c g^2 (1- .60/{N_c}^2) \]
Since the theory is super-renormalizable  it is  plausible that, unlike in four dimensions, the lattice value of $g^2$ and the continuum value $g_s^2$ describing the same physics are not that different.  We therefore assume this relation also for the continuum theory and take $\alpha_s = g_s^2/{2 \pi}$. Finally, we adapt to two space dimensions the standard bag model argument to relate the bag constant,$\Lambda$, the string tension (energy per unit length of the ``flux tube'' between fundamental source and its anti-source) and $\alpha_s$: For general ``tube radius'' $R_t$, we have the energy per unit length of the ``tube'' of flux between fundamental source and anti-source:
\[  \Lambda (2R_t) + \frac{1}{2}{\bf E}^2 (2 R_t),  \]
 where ${\bf E}$ is the color electric field assumed constant across the diameter of the confined flux tube.  Using Gauss's theorem with the color flux as $E (2R_t)$ to express this as a function of the variable $R_t$ only, and minimizing with respect to $R_t$ gives:
\[ R_t = \left(\frac{c_N g_s^2}{8 \Lambda}\right)^{\frac{1}{2}}  \]
\[ \sigma = \sqrt{4 \pi c_{N_c} \alpha_s \Lambda}  \] 
where $c_{N_c} = ({N_c}^2-1)/2 N_c)$ is the SU($N_c$) Casimir in the fundamental representation.

This last relation gives
\[ \Lambda = \frac{\sigma^2}{2 c_{N_c} g^2}   \]
 and we find approximately
\[ \Lambda \approx 0.2 \sigma^{\frac{3}{2}} \left(1 + \frac{0.40}{N_c^{2}}\right)  \]
Thus the expression (3) gives the energies as

\begin{equation}  \mu \beta^{\frac{2}{3}} \sigma^{\frac{1}{2}}\left( 1 + \frac{.13}{N_c^2}\right)  \end{equation}
with $\mu \approx 1.6$.
This gives  the  masses of the single gluon states in two space dimensions given in the third column of Table I. This column neglects  self-energies and the color Coulomb interaction with the source.  Masses are expressed in units of $\sqrt{\sigma}$.
We neglect  a small positive correction of about 3 per cent for $N_c=2$ and smaller for larger $N_c$.
  
Self-energies and the color Coulomb interaction of the gluon and the adjoint source at the origin alter these values. These effects turn out to be small,   so we shall first neglect their influence, by way of  energy minimization, on the bag radius, $R_0$. The dependence of the Coulomb energy on $R_0$ comes from $C$ (see Section IIIB) which may be rewritten as
\[ C = \frac{1}{2} \ln R_0/\beta = -\frac{1}{3} \ln \beta + const = C^{\prime} + const. \]
We give in Table II the values of $A$, $B$ and the total $T=A-B+ C^{\prime}$.\\ 
Taking $\alpha_s N_c$ , in units of $\sigma$,   to be about $0.8\sqrt{\sigma}$  our masses of the lowest one gluon states, corrected for the color Coulomb interaction as a small perturbation, are as given in the second column of Table I.\\  
 We see that the color Coulomb perturbation has not changed the ordering  of the one gluon states. Its effect on states of two and more gluons  is also small. We have also neglected transitions between one- and two-gluon  states inside the bag. We expect these effects to be even smaller than color Coulomb interactions except in the case of degenerate or near degenerate states of the same quantum numbers. 
\section{Models of Glueballs}
We next consider glueballs ie systems with no color source, first in the bag model in a simple approximation and then in the flux-tube model. In each case we shall briefly refer to the corresponding model of gluelumps. In the case of the bag model this will show how the extra approximation of this section affects the spectrum as discussed previously without the simple approximation.

\subsection{Bag Model: Harmonic oscillator approximation to the bag model}
The spectrum of a single gluon in the bag model in two space dimensions ($\beta$ versus $|m|$) has an approximate resemblance to the spectrum of a two dimensional harmonic oscillator (energy versus $|m|$). In the bag model of pure glue states one considers higher states which contain two or more gluons in the fixed bag although for pure pure glue states there is no fixed source defining the bag position. A major difficulty with these states is that many of them are ``spurious''. In other words they are not genuine internal excitations of the system. This difficulty is easy to control for a harmonic oscillator system where the internal modes and centre of mass motion decouple. Therefore we consider instead a detailed two dimensional harmonic oscillator model to mimic the internal modes of the bag model.\\
The spectrum of the two dimensional harmonic oscillator is very simple: the ground state with $m=0$ is at $(\hbar) \omega$, the lowest states of $m= \pm1$ at  $2(\hbar) \omega$, the first excited state with $m=0$ at $3 \omega$, which is also the position of the lowest states with $m = \pm2$ , etc. At each principal quantum number $n$ and energy $(n +1)\omega$ one has a set of states with $m=0,\pm2, \pm4,...$ for $n$ even or $m=\pm1, \pm3, \pm5,...$ for $n$ odd. (See Fig. 1.) The values of $\beta$ of Section IIIA are roughly equally spaced by $\omega \approx 1.4$ with values of $\beta$ displaced upwards by $\kappa \approx 0.7 \omega$ from their harmonic oscillator values. We interpret $\kappa$  as a constant added to the harmonic oscillator ``gluon potential''. The values of $\beta$ determined in this way may then be summed and related to masses by the bag model relation (4), so that effectively we are taking the Hamiltonian to  be proportional to the two-thirds power of a   harmonic oscillator with the harmonic oscillator energy shifted by a constant, $\kappa$.
 
 With \underline{two particles} in the harmonic oscillator well the ground state of the system is at $2 \omega$, corresponding to both particles in the harmonic oscillator ground state, and other states are obtained by putting each particle in a specific state of the single particle oscillator. The resulting states may be separated into excitations of the internal degrees of freedom of the two particle system and centre of mass excitations by introducing appropriate internal coordinates ${\bf r_1-r_2}$ and center of mass coordinates ${\bf R=(r_1+r_2)}/2$. With this simple choice of coordinates, the \underline{internal} states of the two particle system are exactly the same as those of a single two dimensional oscillator with ground state ($m=0$) at $\omega$, first excited states ($m= \pm1$) at $2\omega$ etc. This spectrum is identical to that of a single particle in the two dimensional oscillator. If we interpret this spectrum as giving the $\beta$ values in Eqution (4) of a system   of two confined ``gluons'', we   note that the physical states are color singlets which must be symmetric in color space (${\bf F_1 \cdot F_2}$), and only spatial states which are symmetric under interchange of the two ``gluons'' are allowed by Bose symmetry. This rules out states with $m$ odd which are antisymmetric under the interchange $1\leftrightarrow2$. So the only physical states for two ``gluons'' in this model have $m=0,\pm2,\pm4,...$. Therefore for two ``gluons'' we have a ground state at $\omega$, an excited ($m=0$) state at $3 \omega$ and an excited $m=\pm2$ doublet at $3\omega$. All these states have positive charge conjugation. There are higher states at $5\omega,7\omega,...$ with similar quantum numbers ($|m|=0, 2,4,6,8,.., C=+1$).\\
For three ``gluons'' the harmonic oscillator analogue contains three particles in a two dimensional oscillator. After eliminating centre of mass motion there are two   internal coordinates, say \boldmath ${\rho}$ \unboldmath $={\bf (r_1-r_2)}/\sqrt{2}$, \boldmath$\lambda$\unboldmath $= {\bf(r_1 + r_2} -2 {\bf r_3})/\sqrt6 $  
which are two dimensional vectors. The spatial wave functions of the three ``gluon'' system contain homogeneous polynomials in \boldmath$ \rho, \lambda$ \unboldmath with definite angular and parity behaviour and definite permutation properties under the interchange $(12)$ and $(13)$. The color wave functions of the three ``gluon'' system in a singlet state are either totally symmetric $d_{ijk} F^{(1)}_i F^{(2)}_j F^{(3)}_k$ or totally antisymmetric $f_{ijk} F^{(1)}_i F^{(2)}_j F^{(3)}_k$. The three ``gluon'' state with $f_{ijk}$ coupling has positive charge conjugation, while the $d_{ijk}$ coupling has negative charge conjugation. Therefore Bose symmetry for three ``gluons'' allows only totally symmetric ($S$) spatial wave functions for $C=-1$ states and totally antisymmetric spatial wave functions for $C=+1$ states. All other spatial states, which have mixed symmetry, do not correspond to physical (color singlet) glueballs. The lowest internal state is at $2 \omega$ with an ${\sf S}$ wave function, and has quantum numbers $|m|^{PC} = 0^{--}$. The parity $P=-1$ because of the intrinsic parity of each ``gluon''. The next allowed color singlet three ``gluon'' state is an excitation corresponding to the polynomial \boldmath ${\bf\rho^2 + \lambda^2}$ \unboldmath and is also $0^{--}$, at energy $4 \omega$. Degenerate with it is a multiplet with $m=\pm2$ and $C=-1$ which has  polynomial $(\rho_{\pm}\rho_{\pm} + \lambda_{\pm}\lambda_{\pm})$ where $\rho_{\pm} = \rho_x \pm i\rho_y$, etc. At one unit of $\omega$ higher, $E=5 \omega$ one can construct (cubic) three ``gluon'' states of $m=\pm 1, \pm 3$, eg \boldmath $({\bf \rho^2 - \lambda^2})$\unboldmath ${\it\lambda_{+}} + 2$\boldmath${\bf (\rho . \lambda)}$\unboldmath ${\it\rho_{+}}$   being a totally symmetric state of $m=+1$, and therefore $C=-1$, corresponding to a multiplet with $|m|^C = 1^-$. One can construct similarly, at $5 \omega$, multiplets with $|m|^C = 1^+, 3^-, 3^+$. Thus we have found the lowest states of $|m|^C= 1^+,1^-$, higher in energy than the lowest states of $|m|^C= 2^+,2^-$.   At $E = 6 \omega$ one can find states with $|m|^C =0^{--}, 0^{++}, 2^+, 2^-, 4^-$ for the three ``gluon'' system. At $7 \omega$ there are states with $1^+, 1^-,3^+, 3^-, 5^+, 5^-$. One can also construct   states having quantum numbers   $0^{-+}$ and $0^{+-}$ with three ``gluons'', but these are rather high, at $E = 8 \omega$ and $E = 10 \omega$ respectively. For example the state with $0^{-+}$ has totally antisymmetric wave function of type $(r_1^2-r_2^2)(r_2^2-r_3^2)(r_3^2-r_1^2)$ which is six units of $\omega$ above the ground state at $2 \omega$. However these   quantum numbers appear at lower energies in the spectrum of four ``gluon'' states.
\\With four ``gluons'' there are eight independent ways of constructing a color singlet state in SU(3). Of these eight states three have negative charge conjugation. The three color wave functions of $C=-1$ transform like the anti-triplet ${\bf \bar{3}}$ representation of the permutation group of four objects, ${\sf S}_4$. Therefore, to obey Bose statistics, a $C=-1$ states should have orbital wavefunction of the same  ${\bf \bar{3}}$ symmetry.  To construct the lowest $0^{+-}$ state we require the wave function to be built from polynomials in \boldmath $\rho^2$, $\lambda^2$, $\sigma^2$, ${\bf \rho . \lambda}$, ${\bf \rho . \sigma}$ and ${\bf \lambda . \sigma}$, where ${\bf \rho}$ and ${\bf \lambda}$ are as in the three ``gluon'' case and ${\bf \sigma}$\unboldmath$={\bf (r_1 + r_2 +r_3} - 3 {\bf r_4})/\sqrt{12}$. The coordinates \boldmath${\bf \rho}$, ${\bf \lambda}$ and ${\bf \sigma}$ \unboldmath transform like a triplet ${\bf 3}$ representation of ${\sf S}_4$. The scalar products \boldmath $\rho^2$, $\lambda^2$,...,${\bf \lambda . \sigma}$ \unboldmath transform like ${\bf 1 +2 +3}$ of ${\sf S}_4$. The lowest polynomial which transforms like ${\bf \bar{3}}$ of ${\sf S}_4$ are quadratics in \boldmath$\rho^2$, $\lambda^2$,...${\bf \lambda . \sigma}$ \unboldmath and occur in the Kronecker product ${\bf 3 \otimes 2}$. Therefore the lowest $0^{+-}$ four ``gluon'' state (in SU(3)) is at $4\omega$ above the ground state of the four ``gluon'' system ie at $E = 7\omega$, somewhat lower than in the three ``gluon'' sector (where it was at $10\omega$).  The other set of quantum numbers  whose lowest state lay quite high in the three ``gluon'' sector was $0^{-+}$. This requires a color wave function of positive charge conjugation.  Of the five color states with $C=+1$, one is totally symmetric under permutations and the other four form two doublets. Therefore the spatial wave function of a $0^{-+}$ state should be either totally symmetric or part of an ${\sf S}_4$ doublet. To obtain negative parity one needs a factor of \boldmath ${\bf \rho \times \lambda}$, ${\bf \lambda \times \sigma}$ or ${\bf \sigma \times \rho}$ in the wavefunction. These forms transform under permutations like a ${\bf \bar{3}}$ representation of ${\sf S}_4$. In order to satisfy Bose statistics we need to construct a doublet representation. This we can do with a fourth  order polynomial by taking  a product of the ${\bf \bar{3}}$, (${\bf \rho \times \lambda}$) etc with a ${\bf 3}$ constructed from the scalars $\rho^2$, $\lambda^2$,...,${\bf \lambda . \sigma}$ \unboldmath. Therefore the lowest $0^{-+}$ state with four ``gluons'' has $E = 3\omega + 4\omega= 7\omega$ in this model. Therefore we find that the lowest four ``gluon'' states of quantum numbers $0^{+-}$ and $0^{-+}$ are degenerate at $E = 7 \omega$, and  these appear to be the lowest states of these quantum numbers for any number of ``gluons''.\\  
The spectrum of states for two, three and four ``gluons'', is summarized in Fig. 2. The notation is $|m|_n^{PC}$ where $n$ is the number of ``gluons'' and the $P$ quantum number is of course omitted for $|m| \neq 0$. The lowest lying states are included for all quantum numbers up to $|m|=6$ though the list of higher excited states with $n \geq4 $ is not complete.
\\We could have used the harmonic oscillator approximation also in calculating the bag model gluelump spectrum, but this extra approximation is not necessary there as there is no problem of spurious degrees of freedom in that case. The effect would have been to make exactly degenerate states which in the calculation of Section III were slightly separated in energy.

\vspace{0.3cm}    
\subsection{Flux-tube model of two dimensional glueballs}
The flux-tube model \cite{IP} was invented as an opposite extreme to the bag model. The bag model uses essentially the same degrees of freedom, quarks and gluons, as perturbation theory. The flux-tube model, on the other hand, postulates that in the confining regime the appropriate gluonic degrees of freedom are the flux links of lattice Hamiltonian gauge theory \cite{Kogut}. In this model the simplest pure glue states consist of a loop of flux whose quantum dynamics gives rise to the spectrum of states. Ref \cite{IP} suggested that a non-relativistic string Hamiltonian might capture the essence of the flux-tube dynamics. In the case of three space dimensions, a further adiabatic assumption was required to make the system tractable.
\\ The corresponding model applied to two space dimensions was considered in \cite{Moretto} for the case of color SU(2) and a generalization for more than two colors was given in   \cite{Johnson}. In each case, the spatial configuration of the flux-tube is assumed to be expressible  in the functional form 
 $r = r(\phi)$  in polar variables $(r,\phi)$ where $r(\phi)$ is expressed in a Fourier series
\[ r = {\sum}_{n=-\infty}^{n=\infty} r_n \exp(in\phi) \]
For SU(2) the corresponding non-relativistic Hamiltonian is
\begin{equation}
 H = 2\pi \sigma r_0 + \frac{1}{4\pi r_0  \sigma}p_{r_0}^2 + \frac{1}{r_0} H_{osc} \end{equation}
with
\begin{equation}
H_{{\rm osc}} =  \sum_{n \neq 0, \pm 1}|n|a_n^+a_n +c \end{equation}
where $H_{{\rm osc}}/r_0$ is the Hamiltonian for deviation of the flux-tube shape (assumed small) from a circle, with $a_n^+$ and $a_n$ creating  and annihilating quanta of the corresponding $n$th vibration mode of the flux-tube and $c$ arising from the (renormalized) zero point energy of the oscillators. Note that, in the summation over $n$, the values $n= \pm1$ may be  excluded since these correspond simply to a translation of the flux-tube without changing its shape. It follows that there are no spurious centre of mass excitations in the flux-tube model. Note also that $H_{{\rm osc}}$ depends on variables which are independent of $r_0$ and $p_0$ and therefore commutes with them. Variables may therefore be chosen to diagonalize $H_{{\rm osc}}$ for all $r_0, p_{r_0}$, and the corresponding eigenvalues are
\[{\sf N} = \sum(|n| l_n) +c  \]
where $l_n$ is the degree of excitation of the $n$th mode.
The difference between  the case of two colors and those of more than two colors is, in the flux tube model, the difference between whether the flux-tube has or does not have an orientation arrow along its length. To the flux-tube shape ${\sf C}: r = r(\phi)$ there  corresponds   the non-local gauge invariant path ordered trace
\[ Tr_{\sf P} \exp (i \int_{\sf C}\bf{A}.d{\bf r})  \]
where ${\bf A}$ is the color vector potential operator ${\bf A}_i \lambda_i$ and ${\sf P}$ orders matrices in the expression around the path ${\sf C}$. This operator acting on the (strong coupled) vacuum creates a line of color electric flux (in the fundamental representation) around ${\sf P}$ \cite{Kogut}. States of the flux-tube model are obtained by   summing over paths ${\sf P}$  with appropriate weightings \cite{IP}. In color SU(2), where the $\lambda$s are just Pauli matrices, this trace is identical to the trace with all matrices replaced by their complex conjugates, because in SU(2) the fundamental representation is isomorphic to its conjugate.  Therefore the trace operator necessarily has charge conjugation $+1$ and so all states of the flux tube model will have charge conjugation $+1$. It also follows that the trace over 
 ${\sf C}$ is identical to that over the reversed path ${\sf C'}$. This is what we mean by saying that  the flux tube has no orientation. 
\\For higher color groups, on the other hand, the path ordered trace is {\it not} identical to that with all matrices replaced by their conjugates and the operators for the paths ${\sf C}$ and ${\sf C'}$ are distinct. Thus both $C=+1$ and $C=-1$ states are possible in the case of more than two colors. In \cite{Johnson} the identical Hamiltonian is assumed to describe flux-tubes of both orientations and states of charge conjugation $+1$ or $-1$ are necessarily degenerate, corresponding to the sum and difference of flux-tube states of opposite orientation. Formally this is equivalent to saying that the positive and negative orientation states live in different spaces  connected by the charge conjugation operator.
\\The form of the flux-tube model described so far assumes
\begin{enumerate}
\item non-relativistic quantum mechanics, as stated above
\item the coefficents $r_n$ for $n \neq 0$ are small enough so that the flux-tube shape exhibits only small deviations from circular. This is required both because a single-valued function $r(\phi)$ cannot describe all flux-tube shapes (eg those which are self-intersecting) and also the harmonic approximation to the string potential energy is only valid in this case.
\end{enumerate}
The first of these assumptions may plausibly be circumvented by noting that the non-relativistic Hamiltonian above may be regarded as the first two terms in an expansion in powers of $1/r_0$ of $H_{rel} = \sqrt{M^2}$ with 
\begin{equation} M^2 = (2\pi\sigma r_0)^2 + p_{r_0}^2 + 4\pi\sigma H_{osc}
\end{equation}
We observe that in this last relativized form the mass squared operator of the flux-tube is simply that of an infinite set of harmonic oscillators, one for radial motion and the rest for deviations from the circular form. Eigenstates will be characterized by the set of integer excitation numbers $\{ l_n \}$ with corresponding energies $$4\pi\sigma (N +k)= 4 \pi \sigma \left(l_0 -1 + {\sum}_{n\neq0} l_n |n| +k\right)$$ where $k$ is a new unknown (renormalized) zero point energy directly related to the mass of the ground state glueball. For positive and negative $n$, $l_n$ may be zero or a positive integer. For the case of $n=0$ the corresponding radial excitation number $l_0$ must be odd, and, as stated above, $n=\pm1$ are excluded from the sum. The oddness of $l_0$  is required to make the non-relativistic Hamiltonian or the squared mass operator Hermitian, since the radial variable $r_0$ must be positive and hence the wave function must vanish at $r_0=0$.
\\The angular momentum quantum number of the states is also easy to work out \cite{Moretto}. The operator for exciting one quantum of the $n$th mode has angular momentum $n$. Thus the states with no excitations other than excitation of the radial mode $n=0$ have $m=0$; those with one excitation of mode $n=\pm2,\pm3...$ have angular momentum quantum number $m=n$ and for multiple excitations the contribution to the  angular momentum quantum number are additive.
\\Starting with the case of color SU(2) \cite{Moretto} we immediately deduce the following features of the two plus one dimensional glueball spectrum in the flux-tube model as we have defined it above:
\begin{enumerate}
\item The mass of a state depends on $N = l_0 -1 + \sum(|n| l_n)$ only, with the ground state having $N=0, l_0=1$. The squared mass  is a linear function $M^2 = 4\pi\sigma(N + k)$ of $N$.
\item There are sequences of radial excitations, $l_0=3, 5, 7...$equally spaced in squared mass and separated by $8 \pi \sigma$.
\item the lowest states of $m=2$, which have one $|n|=2$ quantum, have squared mass $8 \pi \sigma$ above the ground state ie degenerate with the first radial excitation. For larger $|m| = 3,4..$ the lowest state differs from the ground state by $4 \pi \sigma |m|$ in squared mass.
\item The lowest mass states are in order
\[ N = 0:\;\;\;|m|^P = 0^+ \]
\[ N = 2:\;\;\; |m|^P = 0^+, |m| = 2 \]
\[ N = 3:\;\;\;\;\;\;\;\;\;\;|m| = 3  \] 
\[ N = 4:\;\;\; |m|^P = 0^+,  |m| = 2, |m| = 4)  \]
\[ N = 5:\;\;\; |m| =1,  |m| = 5, |m| = 5  \]  
The lowest state of quantum numbers $0^-$ has $N = 8$ and is a linear combination of states with  non-zero excitation numbers, $l_{|2|} = 2$, $l_{|4|} = 1$. It lies in energy well above those listed above. (All these states have of course $C=+1$ as required by the general theory.)  
\end{enumerate}
The spectrum of the SU(2) flux-tube model is given as the $C= +1$ states of Fig 3, which also incorporates the results for the spectrum of flux-tube Model II for more than two colors (see below).  The notation is $(r^{n_0-1}\Pi_{n\neq0,\pm 1}{|n|^{\it sign(n)} })^{PC}$ with $P$ present only for $m=0$. 

\vspace{0.3cm}  
Now going over to the case of more than two colors, in the relativized version of the model of \cite{Johnson} which we call Model I, the radial and angular oscillators are all doubled , one corresponding to each of the two orientations of ``glue loop'', and the mass eigenstates are also doubled to form degenerate $C=\pm1$ states at each $N$.  In   particular the SU(2) states of $(m^P) = (0^ +)$ correspond to a degenerate pair $(m^{PC}) = (0^{++}), (0^{- -})$. Thus for this model of more than two colors, the level ordering is
\[ N = 0:\;\;\;|m|^{PC} = 0^{++}, 0^{--} \]
\[ N = 2:\;\;\;m^{PC} = 0^{++}, 0^{--}, |m|^C = 2^\pm \]
\[ N = 3:\;\;\;\;\;\;\;\;\;\;|m|^C = 3^\pm  \] 
\[ N = 4:\;\;\;m^{PC} = 0^{++},0^{--}, |m|^C = 2^\pm,|m|^C = 4^\pm  \]
\[ N = 5:\;\;\;|m|^C =1^\pm,  |m|^C = 5^\pm, |m|^C = 5^\pm  \]
The doubling of all oscillators as well as the $C$ degeneracy of all levels seems somewhat profligate, and perhaps unphysical. In particular it would appear more natural if the same angular oscillator operators could be applied to a string of either orientation. We may then get away with doubling the radial oscillators only. If, as seems very plausible, we take the lowest state to be non-degenerate and having $C=+1$, the creation operator formed from one   radial oscillator, variables $r_1, p_{r_1}$ will produce clockwise excitations of the system and that from the other radial oscillator with variables   $r_2, p_{r_2}$ will produce counter-clockwise  excitations. We call this new model for more than two colors Model II. The mass squared operator of Model II is  
\begin{equation} M^2 = (2\pi\sigma)^2 (r_1^2 + r_2^2) + p_{r_1}^2   + p_{r_2}^2 + 4\pi\sigma H_{osc}
\end{equation}
To obtain a $C=-1$ state in this case we must form a state antisymmetric in $r_1, r_2$. Thus there is no $C$-degeneracy of the states unless there is some excitation in the radial coordinate. The spectrum of Model II will be identical to that of the two color model with the exception that
\begin{enumerate}
\item The first radial excitation of any $|m|$ state will now house states of both $C=+1$  and $C=-1$ the latter containing the factor $r_1^2 - r_2^2$.
\item The second radial excitation of any $|m|$ state will hold one $C=-1$ state with wave function containing an extra factor antisymmetric in $r_1^2$ and $r_2^2$ and two $C=+1$ states containing an extra factor  symmetric in $r_1^2$ and $r_2^2$.
\item The third radial excitation will hold two $C=1$ and two $C=-1$ states etc.
\end{enumerate}
The low lying spectrum of Model II is shown in Fig 3.\\  
To conclude this section on the flux-tube model of glueballs we emphasize again that our description of the flux-tube in terms of polar coordinates is to some extent an unphysical restriction since it excludes reentrant curves. This should not be a serious deficiency for   states highly excited in the radial variables but is liable to be serious especially for the ground states of any set of quantum numbers. It is this feature that has forced  the degeneracy in $C$ of Model I   since curves given by single valued $r = r(\phi)$ are necessarily oriented. The second model with the two radial oscillators goes beyond this by postulating a non-degenerate non-oriented ground state. Of course, classically the ``glue loop'' in the ground state just reduces to a point. Furthermore ,   the states of maximum $|m|$  produced by exciting only the $n = \pm 2$ oscillators correspond classically to the limiting motions in which the loop lies on a rotating straight line. The quantum states  are held from collapsing by quantum fluctuations. It is  plausible that, in a larger space of loops, because of these quantum fluctuations the  orientations of these states are lost. Another possible model would be simply to take the spectrum of Model I but with the removal of the   $C = -1$ states corresponding to these limiting classical motions. It differs only slightly from Model II. (eg It would unlike Model II give a $C= -1, |m|=3$ state at $N=3$ and the degeneracies of some radially excited levels would be different.)
   
\subsection{Extension of flux-tube model to two dimensional gluelumps}
Having treated above the case of glueballs in the flux-tube model (``glue loops''), the case of gluelumps is a straightforward generalization. We assume that the wave functionals of gluelumps are given by  linear combinations of loop traces where the loop goes through the origin and the quantity in the trace contains an extra factor of a $\lambda$ matrix corresponding to the source at the origin. We note the following:
\begin{enumerate}
\item For two colors the trace is necessarily odd (even) under the replacement of Pauli matrices $\lambda_1$, $\lambda_3$ ($\lambda_2$) by their conjugates, ie $C$ is necessarily the same as for a single gluon field (called C=-1; see Section II). For more than two colors both values of $C$ are possible. 
\item We  take the mass squared operator for the case of two colors to be given in terms of a parametrization of the loop through the origin as a distorted circle with its centre at $(x,y)$. The polar coordinate $r$ of a point on the loop is defined as the  distance from  the centre $(x,y)$ and the polar angle $ \theta$ is measured from the direction defined by the vector $(x,y)$.  We have
\[ r(\theta) = \sqrt{x^2 + y^2} + {\sum}_{n=2}^{\infty} r_n \sin( n \theta)  \]
where the Fourier expansion contains  sines only to ensure that the curve goes through the origin. The $n=1$ term is omitted since it corresponds to rotation about the origin which is also effected by a transformation of the variables $(x,y)$. 
\item The relativised squared mass operator for two colors  analogous to equation (7) is 
 \begin{equation} M^2 = (2\pi\sigma)^2(x^2 + y^2) + p_x^2 + p_y^2 + 4\pi\sigma H^{\prime}_{osc}
\end{equation}
where 
\[ H^{\prime}_{osc} =\sum _{n=2}^{\infty} n b_n^+ b_n \]
and $b_n$, $b_n^+$ are the annihilation and creation operators for sinusoidal deviations from the circular shape. Apart from an arbitrary zero point contribution the eigenvalues of $M^2$ are again $4\pi\sigma N$ where $N$ is an integer.
\item The corresponding spectrum for two colors is:
\begin{itemize}
\item the spectrum of the $(x,y)$ two dimensional oscillator (recall  Fig 1): 
\begin{itemize}
\item ground state of $m^{PC}=0^{--}$ with $N=0$ 
\item  radial excitations of the ground state separated from it in mass squared by even multiples $N$   of $4\pi\sigma$
\item the lowest lying states of $|m| =1,2,3,...$ and negative $C$ separated from the ground state by    $4\pi\sigma N$ with $N=1,2,3,..$, and their radial excitations again even multiples of $4\pi\sigma$ above these.
\end{itemize}
\item states obtained from any of these two dimensional oscillator states by excitation by an arbitrary number of units in the oscillators of $H_{{\rm osc}}$, with further vibrational contribution to the squared mass given by $4\pi\sigma {\sum}_{n =2}^{\infty} i_n n$ where the $i_n$ are positive integers or zero. In this case, unlike the case of glueballs in the flux-tube model, the  angular momentum quantum number is not affected by the excitation of the loop shape. The vibrational contribution to the parity of an $|m|=0$ state is $(-1)^{1+\sum i_n}$ 
\item the lowest state of $m^{PC}=0^{+-}$ has $N=2$. The quantum numbers $m^{PC}=0^{-+}$ do not occur till $N=4$.
\end{itemize}
\item For more than two colors, we may modify the model for gluelumps to be analogous to either of the two corresponding glueball models I and II discussed above. As for glueballs, Model I will simply double all states introducing  $C=+1$ partners for all states. The analogue of the more physical Model II   involves two radial variables $r_1$ and $r_2$ (rather the one radial variable $r=\sqrt{x^2 + y^2}$ of the SU(2) case) and introduces extra states of opposite $C$    only at the level of the first  and higher radial excitations.\\
\end{enumerate} 
 
To summarize, the extension of the flux-tube model to two dimensional gluelumps gives the lowest states  having squared mass $4\pi\sigma(N+ {k}^\prime )$ with quantum numbers as given in Table III.\\  
     
\section{Comparison with lattice calculation and conclusions}
The model calculations presented above can obviously not be compared to experiment, but in the case of glueballs in two space dimensions extensive lattice Monte Carlo calculations are published \cite{Teper} for $N_c=2,3,4,5$ colors, and in this section we first examine how the two dimensional glueball spectra of the bag model and flux-tube model   compare with each other and with the lattice results. For convenience we reproduce Teper's color SU(2) and SU(3) results in Table IV (columns 2 and 3). The results for the higher color groups are similar to SU(3) though Teper has attempted to extract a dependence on $N_c$.\\
We first note that we do not expect either model or indeed the lattice to give a completely accurate picture of the spectrum. For one thing, it is clear on general grounds that in the full two plus one dimensional theory only a few of the states of lowest mass can be stable, higher states being able to decay. Thus the infinite tower of stable states produced by each model is at best an approximation. There is also the question of whether the higher states as determined on the lattice correctly incorporate decay channels and the shifts in mass that they induce, since on the lattice the momenta of final decay products  are forced to be discrete. In addition there is the problem   that on a square lattice the quantum number $m$ is determined in a simple way only modulo $4$. This may introduce substantial real ambiguity into the present lattice results.(A method of resolving it is suggested in the second item of reference \cite{Johnson}.) We also remind the reader that each of our models has its own limitations which must necessarily restrict its accuracy even if it is basically correct physically: 
 
\begin{itemize}
\item We had to simplify the bag model substantially in order to solve in a simple way the problem of spurious centre of mass excitations. This we did through our oscillator model of ``gluons''. As the choice of a sharp bag radius  is  also   a simplifying assumption,   we do not feel that replacing this with a harmonic oscillator does any further violence to the physics. The values of the two parameters $\omega$ and $\kappa$ of our bag model spectrum were determined approximately from the gluon modes in the bag (the values of $\beta$ Section IIIA). Fitting the exact modes of Section III to 
\[m=0:\:\:\: \kappa +\omega,  \kappa +3\omega, \kappa +5\omega ... \]
\[m=1:\:\:\:  \kappa +2\omega, \kappa +4\omega ... \]
etc gives  $\omega \approx 1.4$ and $\kappa \approx 0.7\omega$ as quoted in Section IVA.   These choices  are correct to about $\pm .3$ the values of $\beta$ for all of the lowest fourteen modes listed in Section III. An extra ``gluon'' therefore gives an extra contribution to $\beta$ of $\kappa \approx 0.7 \omega$ so that in Fig. 2 states with three ``gluons'' would be expected to be raised by about $0.7 \omega$ compared to two ``gluon'' states and four ``gluon'' states by about twice this amount. The overall constant $\mu$ determined in Section IIIC to be $\approx 1.6$ is also quite uncertain. In the bag model column of Table IV we have kept the suggested values of $\omega$ and $\kappa$ and increased $\mu$ by a factor of $1.2$ so that the bag model ground state of the system agrees with the lattice result for SU(3). This leads to the following formula for glueball masses which we use in the column marked ``bag model'' in Table IV:
\[
M(n^{'},n) = 4.3 ((n^{'} +1 + 0.7 n)/(1 +2(0.7)))^{\frac{2}{3}} 
\]
where $n$ is here the number of ``gluons'' in the bag, and $n^{'}$ is the number of excitations of the oscillator system. For $n^{'} = 0$ and $n = 2$ the formula gives the lattice ground state mass as required, and gives a reasonable fit to higher masses. We emphasize that this formula is not a best fit to the lattice data.
\item The flux-tube picture is expected to be accurate only for states which are physically large enough (compared to the finite resolution necessary to justify the strong coupling picture on which the flux-tube model is based), ie it should be least accurate for the states of lowest mass. This is exactly where our two versions I and II of the flux-tube model for more than two colors differ most. Since Model I exhibits the unphysical charge conjugation degeneracy  of the ground state we shall discard it and use only Model II in our comparisons from now on. This model contains only one parameter, the zero point energy parameter $k$. In contrast to the case of the bag model where an approximate estimate of its overall strength parameter $ \mu$ may be made, we can say very little about $k$. Its value may well be expected to be different in the case of two colors from the case of more than two colors, and   there may  even be further $N_c$ dependence. We ignore these effects as they do not affect the basic ordering of levels and present in Table IV masses for the flux-tube model with $k=1.5$ required again to agree with the lattice  ground state for SU(3). 
\end{itemize}
The first and somewhat surprising conclusion which may be drawn from these comparison is the similarity of the two models certainly at the qualitative level and also the similarity of both models with the lattice data. For example,  
\begin{enumerate}
\item In the $C=+1$ sector, the models both give  the ground   $0^{++}$ state with, at the first excited level, degenerate states of quantum numbers $0^{++}$ and $2^+$. In the lattice data the $0^{+*}$ state is a little lower than the $2^+$ state, but these are certainly the next two $C=+1$ lattice states.
\item The lowest $C=-1$ state of both models has $|m|^P=0^-$. This also agrees with the lattice result. In the flux-tube model it is also degenerate with the first excited $C=+1$ states. In the bag model it appears at $E=2\omega$ in Fig. 2 whereas the first excited $C=+1$ states are at $E=3\omega$. However it is a three ``gluon'' state as opposed to the two ``gluon'' states at $E=3\omega$. This raises $E$ by about $0.7 \omega$, making it just below the degenerate  $0^{++}$ and $2^+$ pair. The lattice results give the lowests $0^{--}$ state degenerate within errors with the first excited $0^{++}$ state, and therefore (see above) a little lower than the lowest $2^+$ state. 
\item Both models agree with the lattice results in giving the perhaps surprising result that the lowest   $|m|=1$ states is higher in mass than the lowest  $|m|=2$ states.
\item Both models also agree with the lattice in giving the lowest  states of quantum numbers $0^{+-}$ and $0^{-+}$ very high. On the lattice the  $0^{-+}$ is at $9.95(32)\sqrt{\sigma}$ in SU(2) and $9.30(25)\sqrt{\sigma}$ in SU(3) and the $0^{+-}$ (which does not exist in SU(2)) at $10.52(28)\sqrt{\sigma}$. These are both more than twice the ground state mass and are above no less than three states of quantum numbers $0^{++}$, at least two states of $2^+$ and (for SU(3)) two states of quantum numbers $2^-$.   
\end{enumerate}
In more detail we observe that with the particular choice $\kappa=\omega$ (\underline{not} the choice $\kappa = 0.7 \omega$ of Table IV) in the harmonic bag model the spectra in $\omega$ of the bag and in $N$ of the flux-tube model are almost identical.  One minor discrepancy between the two models is that in the bag model the $0^{-+}$ and $0^{+-}$ states are degenerate four ``gluon'' states at $E=7\omega$ (ie with $\kappa=\omega$ would correspond $N=8$ in the corresponding flux-tube model),  whereas in the flux-tube model model the lowest $0^{-+}$ state is at $N=8$ and the lowest $0^{+-}$ state does not occur till $N=10$. The lattice data  indeed give the flux-tube model ordering for these two levels with rather large errors, though the  mass values for the flux-tube model as given in Table IV are somewhat higher than the lattice values. Indeed the fact that the $0^{-+}$ state of the flux tube model is too high compared to the lattice was already noted as one of the main discrepancies between the SU(2) flux tube model and the lattice{\cite{Moretto}}. The harmonic bag gives a very similar value for the mass of this state. An interesting discrepancy between the two models is that the flux-tube model gives an expected $3^+$ state at $N=3$ ie corresponding to a mass of around $7.5\sqrt{\sigma}$, whereas the corresponding bag model state is expected somewhat higher, about $9\sqrt{\sigma}$. The lattice data would in this respect somewhat favour the bag model, though the evidence is   indirect. The quantum number $m$ is only easily determined on the lattice modulo $4$, and therefore it would be hard to distinguish $|m|=3$ from $|m|=1$. However, whatever its interpretation in the continuum such a lattice state does not occur till around $10\sqrt{\sigma}$.\\
We do not consider that a detailed fit with either model is called for, since both models are at best qualitative. Such a fit would be driven by the lattice data points with the smallest errors and would therefore not necessarily   reflect the qualitative situation. The sets of model mass values given in Table IV are obtained with the harmonic bag model parameters $\omega=1.4$ and $\kappa=0.7\omega$ as explained above and an overall constant $\mu$ which is a factor $1.2$ larger than the rough estimate of Section IIIC. The flux-tube model fit uses the   additive constant $k=1.5$.
\\
We find the rather close resemblance between the two models somewhat surprising as they are based on very different physics.  
Finally, we point out that the models  predict  a definite   ordering of the gluelump levels in two space dimensions. Though again the models agree in the ordering of the lowest few states, a comparison of  Tables I and III does reveal some discrepancies. A lattice calculation of the gluelump spectrum in $2 + 1$ dimensions might reveal which model is a better description. Note that the gluelump energy spectrum is arbitrary up to an overall additive constant, so only spacings between levels are physically significant.   
  
\acknowledgements{ We thank Michael Teper for encouraging us to investigate the bag model in two dimensions. We thank PPARC and NSERC, Canada for financial support.

\newpage
\begin{table} 
\begin{center} \[
\begin{array}{|c|c|c|c|c|}
\hline
|m|^{C} (m \neq 0) &$one gluon states$&$one gluon states$&$two gluon states$&$three gluon states$\\
$or $ 0^{PC}&$(with color Coulomb)$&$(no color Coulomb)$&$(no color Coulomb) $&$(no color Coulomb)$\\
\hline
0^{--}&2.81,4.86,6.62&2.90, 5.04 &6.27&6.03\\
 \hline
0^{++}&&&4.62, 6.27, 6.47&\\
\hline
0^{+-}&&&6.47&\\
\hline
1^-&3.73,5.71&3.95,5.91&5.47&\\
\hline
1^+&&&5.47&\\
\hline
2^-&4.72&4.80&&\\
\hline
2^+&&&6.27&\\
\hline
3^-&5.55&5.55&&\\
\hline
4^-&6.15&6.23&&\\
\hline
\end{array}\] \end{center} 
\caption{Masses (in units of $(\sigma)^{1/2}$) and quantum numbers of some low-lying gluelumps in the bag model. States of $C=+1$ do not exist for two colors.}\end{table}
%\newpage
\begin{table}[tb]
\begin{center} \[
\begin{array}{|c|r|r|r|r|}
\hline
n&\beta&A&B&T\\
\hline 
0&2.4048&.51905&.33991&-.1133\\
&5.5201&.99185&.64171&-.2193\\
&6.6537&1.33921&.84616&-.2262\\
&11.7915&1.60047&.9935&-.2154\\
\hline
1&3.8317&.69508&.52478&-.2774\\
&7.0156&1.1534&.76387&-.2598\\
&10.1735&1.46508&.93160&-.2397\\
\hline
2&5.1356&1.13029&.68466&-.0997\\
&8.4172&1.4555&.87170&-.1262\\  
&11.6199&1.6968&1.01097&-.1316\\
\hline
3&6.3802&1.4287&.80633&.0047\\
&9.7610&1.6846&.96107&-.0359\\
\hline
4&7.5883&1.6561&.90372&.0770\\
&11.0649&1.8692&1.03664&.0314\\
\hline
\end{array}\] \end{center}  
\caption{The color Coulomb coefficents A,B C and T (see text)}\end{table}
\begin{table}
\[
\begin{array}{|r|c|c|} \hline
N&$states with two colors$&$additional states for Model II$\\
\hline
0&0^{--}&\;\\
\hline
1&1^-&\;\\
\hline
2&0^{--}, 0^{+-}, 2^-&0^{++}\\
\hline
3&1^-,1^-, 3^-&1^+\\
\hline
\end{array} \]
\caption{Spectrum of the lowest gluelumps in the flux-tube model. The squared mass is $4\pi\sigma (N+{k}^\prime)$ where ${k}^\prime$ is a constant.}
\end{table}
\begin{table} 
\begin{center} \[
\begin{array}{|c|l|l|l|l|}
\hline
$state$|m|^{PC}&SU(2)&SU(3)&$bag$(\mu=1.9)&$flux-tube$(k=1.5) \\ \hline
0^{++}&4.718(43)&4.329(41)&4.3&4.3 \\ \hline
0^{++*}&6.83(10)&6.52(9)&6.4&6.6 \\ \hline
0^{++**}&8.15(15)&8.23(17)&7.7&8.3 \\ \hline
0^{--}&&6.48(9)&6.1&6.6 \\ \hline
0^{--*}&&8.15(16)&8.0&8.3 \\ \hline
0^{--**}&&9.81(26)&9.7&9.7 \\ \hline
0^{-+}&9.95(32)&9.30(25)&11.0&10.9 \\ \hline
0^{+-}&&10.52(28)&11.0&12.0 \\ \hline
2^{++},2^{-+}&7.82(14),7.86(14)&7.13(12),7.36(11)&6.4&6.6  \\ \hline
2^{-+*}&&8.80(20)&8.3&8.3 \\ \hline
2^{--}&&8.75(17)&8.0&8.3 \\ \hline
2^{--*},2^{+-*}&&10.31(27),10.51(30)&9.7&9.7 \\ \hline
1^{++},1^{-+}&10.42(34), 11.13(42)&10.22(24),10.19(27)&8.9&9.0 \\ \hline
1^{--},1^{+-}&&9.86(23),10.41(36)&8.9&10.3 \\ \hline
3^+&&&8.9&7.5 \\ \hline
4^+&&&8.3&8.3\\ \hline
\end{array}\] \end{center} 
\caption{Glueball masses in units of $\sqrt{\sigma}$ from the lattice as given by Teper [2] (column 2) by the harmonic bag model  in column 3 and by the flux-tube model  in column 4.}\end{table}
\setlength{\unitlength}{.014in}
\newpage
\begin{figure}
\begin{centering}
\begin{picture} (400,250)
\put(15,20){\line(1,0){300}}
\put(15,20){\line(0,1){200}}
\put(15,20){{\circle*{2}}}
\put(45,20){{\circle*{2}}}
\put(75,20){{\circle*{2}}}
\put(105,20){{\circle*{2}}}
\put(135,20){{\circle*{2}}}
\put(165,20){{\circle*{2}}}
\put(195,20){{\circle*{2}}}
\put(225,20){{\circle*{2}}}
\put(255,20){{\circle*{2}}}
\put(300,5){$|m|$}
%\put(15,5){$0$}
\put(45,5){$0$}
\put(75,5){$1$}
\put(105,5){$2$}
\put(135,5){$3$}
\put(165,5){$4$}
\put(195,5){$5$}
\put(225,5){$6$}
\put(255,5){$7$}
\put(45,40){\circle*{3}}
\put(45,80){\circle*{3}}
\put(45,120){\circle*{3}}
\put(45,160){\circle*{3}}
\put(45,200){\circle*{3}}
\put(75,60){\circle*{3}}
\put(75,100){\circle*{3}}
\put(75,140){\circle*{3}}
\put(75,180){\circle*{3}}
\put(75,100){\circle*{3}}
\put(105,80){\circle*{3}}
\put(105,120){\circle*{3}}
\put(105,160){\circle*{3}}
\put(105,200){\circle*{3}}
\put(135,100){\circle*{3}}
\put(135,140){\circle*{3}}
\put(135,180){\circle*{3}}
\put(165,120){\circle*{3}}
\put(165,160){\circle*{3}}
\put(165,200){\circle*{3}}
\put(195,140){\circle*{3}}
\put(195,180){\circle*{3}}
\put(225,160){\circle*{3}}
\put(225,200){\circle*{3}}
\put(255,180){\circle*{3}}
\put(0,220){$E$}
\put(0,36){$1\omega$-}
\put(0,56){$2\omega$-}
\put(0,76){$3\omega$-}
\put(0,96){$4\omega$-}
\put(0,116){$5\omega$-}
\put(0,136){$6\omega$-}
\put(0,156){$7\omega$-}
\put(0,176){$8\omega$-}
\put(0,196){$9\omega$-}
\end{picture}
\caption[x]{\it Spectrum of two dimensional oscillator (``gluon'')}
\end{centering}
\end {figure}
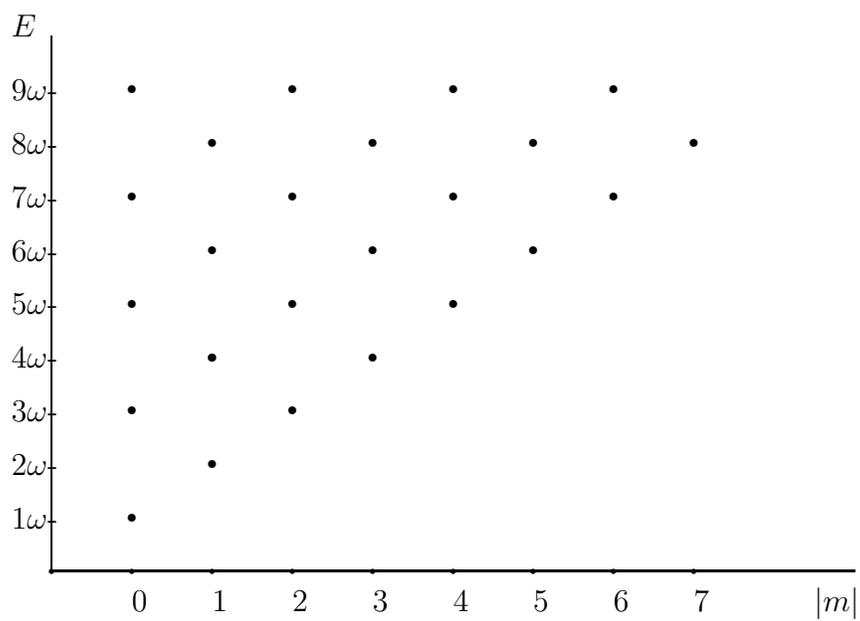
\vspace{0.3cm}
\setlength{\unitlength}{.02in}
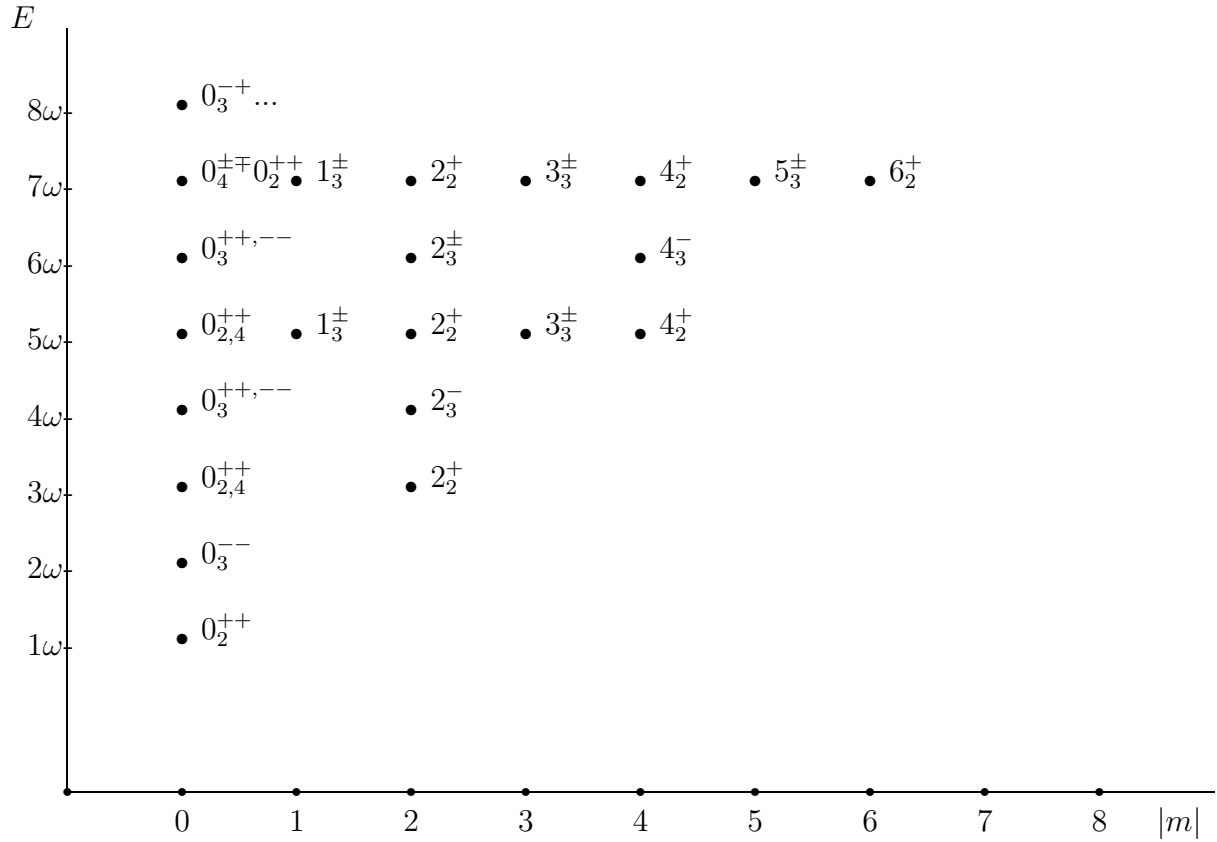
\begin{figure}
\begin{centering}
\begin{picture} (400,250)
\put(15,20){\line(1,0){300}}
\put(15,20){\line(0,1){200}}
\put(15,20){{\circle*{2}}}
\put(45,20){{\circle*{2}}}
\put(75,20){{\circle*{2}}}
\put(105,20){{\circle*{2}}}
\put(135,20){{\circle*{2}}}
\put(165,20){{\circle*{2}}}
\put(195,20){{\circle*{2}}}
\put(225,20){{\circle*{2}}}
\put(255,20){{\circle*{2}}}
\put(285,20){{\circle*{2}}}
\put(300,10){$|m|$}
\put(43,10){$0$}
\put(73,10){$1$}
\put(103,10){$2$}
\put(133,10){$3$}
\put(163,10){$4$}
\put(193,10){$5$}
\put(223,10){$6$}
\put(253,10){$7$}
\put(283,10){$8$}
%\put(45,40){\circle*{3}}
\put(50,60){$0_2^{++}$}
\put(45,60){\circle*{3}}\put(50,80){$0_3^{--}$}
\put(45,80){\circle*{3}}\put(50,100){$0_{2,4}^{++}$}
\put(45,100){\circle*{3}}
\put(45,120){\circle*{3}}\put(50,120){$0_3^{++,--}$}
\put(45,140){\circle*{3}}\put(50,140){$0_{2,4}^{++}$}
\put(45,160){\circle*{3}}\put(50,160){$0_3^{++,--}$}
\put(45,180){\circle*{3}}\put(50,180){$0_4^{\pm\mp}0_2^{++}$}
\put(45,200){\circle*{3}}\put(50,200){$0_3^{-+}...$}
%\put(75,60){\circle*{3}} 
%\put(75,100){\circle*{3}}
\put(75,140){\circle*{3}}\put(80,140){$1_3^{\pm}$}
\put(75,180){\circle*{3}}\put(80,180){$1_3^{\pm}$}
\put(105,100){\circle*{3}}\put(110,100){$2_2^+$}
\put(105,120){\circle*{3}}\put(110,120){$2_3^-$}
\put(105,140){\circle*{3}}\put(110,140){$2_2^+$}
\put(105,160){\circle*{3}}\put(110,160){$2_3^{\pm}$}
\put(140,140){$3_3^{\pm}$}
\put(105,180){\circle*{3}}\put(110,180){$2_2^+$}
\put(140,180){$3_3^{\pm}$}
\put(135,140){\circle*{3}} 
\put(135,180){\circle*{3}}
%\put(135,180){\circle*{3}}
\put(165,140){\circle*{3}}\put(170,140){$4_2^+$}
\put(165,160){\circle*{3}}\put(170,160){$4_3^-$}
\put(165,180){\circle*{3}}\put(170,180){$4_2^+$}

\put(195,180){\circle*{3}}\put(200,180){$5_3^{\pm}$}
%\put(195,160){\circle*{3}}
\put(225,180){\circle*{3}}\put(230,180){$6_2^+$}
%\put(225,200){\circle*{3}}
%\put(285,200){\circle*{3}}\put(290,200){$8_2^+$}
\put(0,220){$E$}
%\put(4.5,36){$1\omega$-}
\put(4.5,56){$1\omega$-}
\put(4.5,76){$2\omega$-}
\put(4.5,96){$3\omega$-}
\put(4.5,116){$4\omega$-}
\put(4.5,136){$5\omega$-}
\put(4.5,156){$6\omega$-}
\put(4.5,176){$7\omega$-}
\put(4.5,196){$8\omega$-}
\end{picture}
\caption[x]{\it Spectrum of glueballs in the  
 harmonic oscillator ``gluon'' model. Mass is proprtional to $\sigma^{1/2}(E + n \kappa)^{2/3}$ where $\kappa$ is a constant and $n$ is the number of ``gluons'' .}
\end{centering}
\end {figure}

\setlength{\unitlength}{.5in}
\begin{figure}
\begin{centering}
\begin{picture} (10,11.5)(.5,-1) 
\put(0,-.5){\line(1,0){10.5}}\put(0,-.5){\line(0,1){11}}
%\put(0,0){\circle*{1}}
\multiput(0,-.5)(2,0){6}{\circle*{.05}}\put(0,-.4){0}
\multiput(0,0)(0,1){9}{\circle*{.05}} 
\put(2,-.4){1} \put(4,-.4){2}
\put(-.4,0){0} \put(6,-.4){3}\put(8,-.4){4}\put(10,-.4){5}\put(10.5,-.4){$|m|$}
\put(-.4,1){1}\put(-.4,2){2}\put(-.4,3){3}\put(-.4,4){4}\put(-.4,5){5}\put(-.4,6){6}\put(-.4,7){7}\put(-.4,8){8}\put(-.4,9){9}\put(-.4,10){10}
\put(-.3,10.6){$N$}
\put(0,0){\circle*{.1}} \put(0,0.2){$(0)^{++}$} 
\put(0,2){\circle*{.1}} \put(.1,2.2){$(r)^{\pm\pm}$} 
\put(0,4){\circle*{.1}} \put(0,4.2){$(r^2)^{\pm\pm}(2^+2^-)^{++}$}
\put(0,6){\circle*{.1}} \put(0,6.2){$(r^3)^{\pm\pm}(3^+3^-)^{\pm \pm}$} \put(0,5.8){$(r,2^+2^-)^{++}$}
\put(0,8){\circle*{.1}} \put(0,8.2){$(4^{\pm}2^{\mp\mp})^{\pm +}$...}
 \put(0,10){\circle*{.1}} \put(0,10.2){$(r,4^{\pm}2^{\mp}2^{\mp})^{\pm\mp}...$}  
\put(2,5){\circle*{.1}} \put(2,5.2){$(2^{\pm}3^{\mp})^{+}$}
\put(2,7){\circle*{.1}} \put(2,7.2){$(r,2^{\pm}3^{\mp})^{\pm}$} 
\put(4,2){\circle*{.1}} \put(4,2.2){$(2^{\pm})^+$} 
\put(4,4){\circle*{.1}} \put(4,4.2){$(r,2^{\pm})^{\pm}$}
\put(4,6){\circle*{.1}}\put(4,6.2){$(r^2,2^{\pm})^{\pm}...$}
\put(6,3){\circle*{.1}}\put(6,3.2){$(3^{\pm})^+$}
\put(6,5){\circle*{.1}}\put(6,5.2){$(r,3^{\pm})^{\pm}$}
\put(8,4){\circle*{.1}}\put(8,4.2){$(4^{\pm})^+,(2^{\pm}2^{\pm})^+$}
\put(8,6){\circle*{.1}}\put(8,6.2){$(r,4^{\pm})^{\pm}$}
\put(10,5){\circle*{.1}}\put(10,5.2){$(5^{\pm})^+,(2^{\pm}3^{\pm})^+$}
\end{picture}
\caption[x]{\it Spectrum of glueball states in the flux-tube model for two colors ($C=+1$ states only) and for more than two colors (Model II). The notation is $(r^{n_0-1}\Pi_{n\neq0,\pm 1}{|l_n|^{\it sign(l_n)} })^{PC}$ with $P$ present only for $m=0$. The squared mass is $4\pi\sigma(N +k)$ where $k$ is a constant.}
\end{centering}
\end{figure}
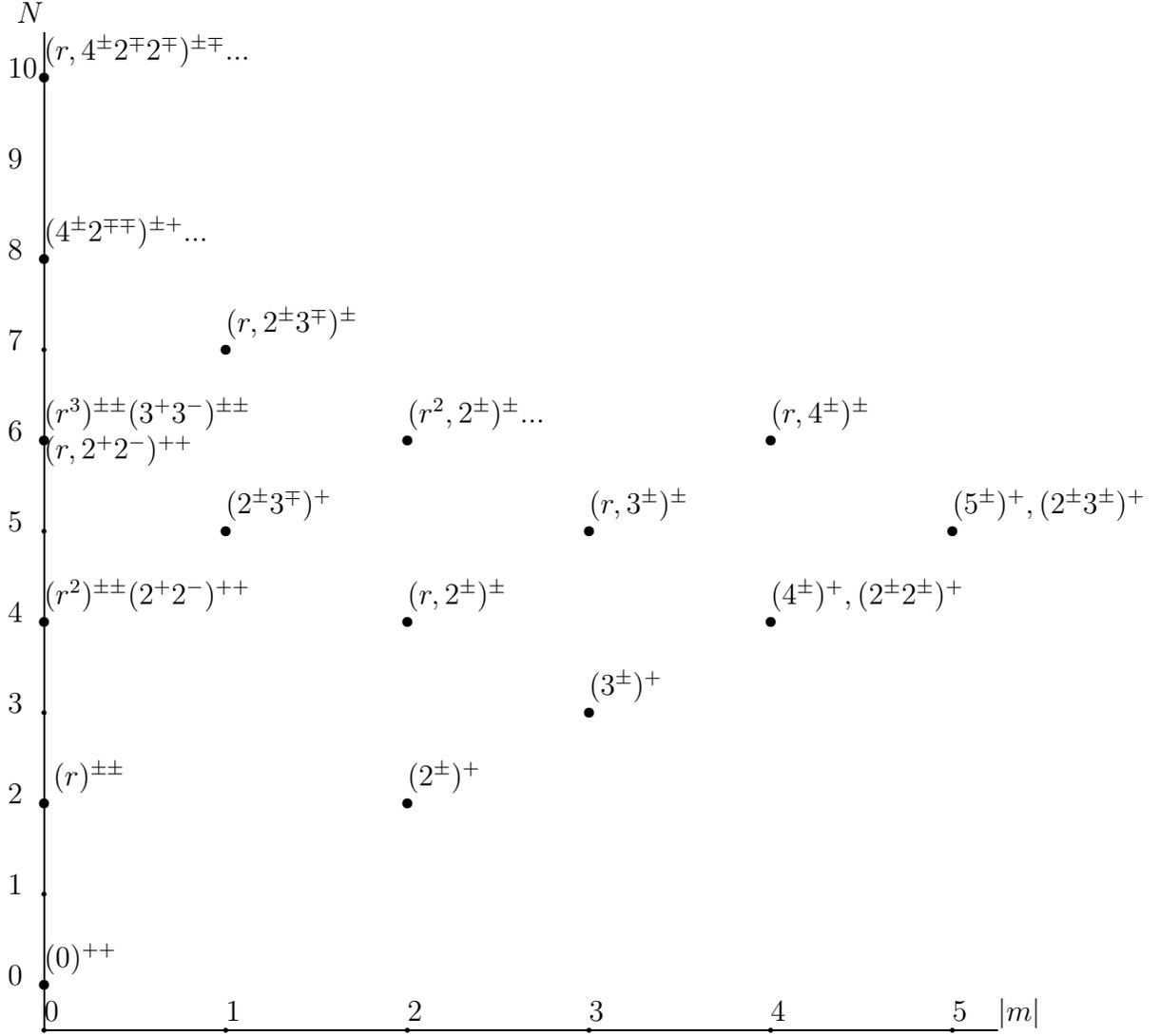 

%\end{flushright}  

\end{document}